\documentstyle[12pt]{article}
\begin{document}
\def\singlespacing{\baselineskip=12pt}
\def\doublespacing{\baselineskip=24pt}

\singlespacing

\pagestyle{empty}

\noindent October 29, 1993 \\
Submitted to Physical Review E \\
cond-mat/9310073
\bigskip

\doublespacing

\begin{center}
\begin{large}
{\bf GAUSSIAN APPROACH FOR PHASE ORDERING IN NONCONSERVED SCALAR SYSTEMS
WITH LONG-RANGE INTERACTIONS}
\end{large}

\bigskip
{\em J. A. N. Filipe and A. J. Bray} \\

\bigskip
Department of Theoretical Physics \\
The University, Manchester M13 9PL \\
\today \\

\bigskip
{\bf  ABSTRACT} \\
\end{center}
We have applied the gaussian auxiliary field method introduced
by Mazenko to a non-conserved scalar system with attractive
long-range interactions. This study provides a test-bed for the
approach and shows some of the difficulties encountered in
constructing a closed theory for the pair correlation function.
The equation obtained for the equal-time two-point correlation function
is studied in the limiting cases of small and large values of the
scaling variable. A Porod regime at short distance and an
asymptotic power-law decay at large distance are obtained. The theory,
is not, however, consistent with the expected growth-law, and
attempts to retrieve the correct growth lead to inconsistencies.
These results indicate a failure of the gaussian assumption
(at least in the form in which we use it) for this system.

\bigskip

\newcommand{\crt}{C({\bf r},t)}
\newcommand{\cRt}{C({\bf r}',t)}
\newcommand{\ck}{C({\bf k},t)}
\newcommand{\pic}{\frac{\pi}{2}C}
\newcommand{\pif}{\frac{\pi}{2}f}
\newcommand{\Bopi}{\frac{2}{\pi}}
\newcommand{\Bpiud}{(2\pi)^d}
\newcommand{\p}{\phi}
\newcommand{\pr}{\phi({\bf r})}
\newcommand{\pR}{\phi({\bf r}')}
\newcommand{\pk}{\phi({\bf k})}
\newcommand{\prt}{\phi ({\bf r},t)}
\newcommand{\mrt}{m({\bf r},t)}
\newcommand{\mA}{m(1)}
\newcommand{\mB}{m(2)}
\newcommand{\SA}{S_0(1)}
\newcommand{\SB}{S_0(2)}
\newcommand{\SAB}{S_0(1)S_0(2)}
\newcommand{\lt}{L(t)}
\newcommand{\ltuB}{\lt^2}
\newcommand{\ltusg}{\lt^{\sg}}
\newcommand{\ltusgA}{\lt^{1+\sg}}
\newcommand{\Dlol}{\frac{\dot{L}}{L}}
\newcommand{\AoluB}{\frac{1}{\lt^2}}
\newcommand{\Aolusg}{\frac{1}{\lt^{\sigma}}}
\newcommand{\jolusg}{\frac{\jlr}{\lt^{\sigma}}}
\newcommand{\jholusg}{\frac{\jlr\,\hds}{\lt^{\sigma}}}
\newcommand{\tuH}{t^{1/2}}
\newcommand{\tuAosg}{t^{1/\sigma}}
\newcommand{\tuAoAPsg}{t^{1/(1+\sigma)}}
\newcommand{\sg}{\sigma}
\newcommand{\ZLsgLB}{0 \!<\! \sg \!<\! 2}
\newcommand{\ZLsgLA}{0 \!<\! \sg \!<\! 1}
\newcommand{\ZLEsgLA}{0 \!\le\! \sg \!<\! 1}
\newcommand{\ZLEsgLB}{0 \!\le\! \sg \!<\! 2}
\newcommand{\ALsgLB}{1 \!<\! \sg \!<\! 2}
\newcommand{\sgLA}{\sg\!<\!1}
\newcommand{\sgGA}{\sg\!>\!1}
\newcommand{\wLLrLLl}{w \!\ll\! r \!\ll\! \lt}
\newcommand{\intd}{\int \! d^d}
\newcommand{\intdk}{\int \! \frac{d^d k}{(2\pi)^d}}
\newcommand{\intdK}{\int \! \frac{d^d k'}{(2\pi)^d}}
\newcommand{\intk}{\int \! \frac{d k}{2\pi}}
\newcommand{\intK}{\int \! \frac{d k'}{2\pi}}
\newcommand{\intdy}{\int \! \frac{d^d y}{(2\pi)^d}}
\newcommand{\intdMA}{\int \! d^{d-1}}
\newcommand{\intinf}{\int_{-\infty}^{+\infty} \!}
\newcommand{\exprk}{e^{i{\bf r}\cdot{\bf k}}}
\newcommand{\expxy}{e^{i{\bf x}\cdot{\bf y}}}
\newcommand{\expmk}{e^{im(1)k+im(2)k'}}
\newcommand{\kdrR}{|{\bf r}-{\bf r}'|^{d+\sigma}}
\newcommand{\kdxX}{|{\bf x}-{\bf x}'|^{d+\sigma}}
\newcommand{\Aoks}{\frac{\jlr}{|s|^{1+\sigma}}}
\newcommand{\hds}{h(d,\sigma)}
\newcommand{\jhds}{\jlr \, h(d,\sg)}
\newcommand{\hdAMs}{h(d,1\!-\!\sigma)}
\newcommand{\hdqMd}{h(d,q\! -\! d)}
\newcommand{\Qds}{Q(d,\sg)}
\newcommand{\qds}{q(d,\sg)}
\newcommand{\bds}{b(d,\sg)}
\newcommand{\btdq}{\beta(d,q)}
\newcommand{\Ads}{A(d,\sg)}
\newcommand{\ads}{a(d,\sg)}
\newcommand{\jads}{\jlr \, a(d,\sigma)}
\newcommand{\jaolusg}{\frac{\jads}{\ltusg}}
\newcommand{\alds}{\al(d,\sg)}
\newcommand{\vp}{V(\p)}
\newcommand{\Dvp}{V'(\p)}
\newcommand{\DvpA}{V'(\p (1))}
\newcommand{\vlr}{V_{LR}}
\newcommand{\Dvlrprt}{V_{LR}'(\phi)}
\newcommand{\DvlrpA}{V_{LR}'(\p (1))}
\newcommand{\jlr}{J_{LR}}
\newcommand{\partprt}{\frac{\partial \phi({\bf  r},t)}{\partial t}}
\newcommand{\partcAB}{\frac{\partial C(1,2)}{\partial t}}
\newcommand{\nabBp}{\nabla^2 \phi}
\newcommand{\nabBcAB}{\nabla^2 C(1,2)}
\newcommand{\DDp}{\frac{d^2\phi(m)}{dm^2}}
\newcommand{\DDpA}{\frac{d^2\phi(m(1))}{dm(1)^2}}
\newcommand{\Half}{\frac{1}{2}}
\newcommand{\dMAox}{\frac{d-1}{x}}
\newcommand{\AoxusgMA}{\frac{1}{x^{\sg-1}}}
\newcommand{\xuAMsg}{x^{1-\sigma}}
\newcommand{\xusgMA}{x^{\sigma-1}}
\newcommand{\AMs}{1-\sg}
\newcommand{\sMA}{\sg-1}
\newcommand{\eq}{\equiv}
\newcommand{\al}{\alpha}
\newcommand{\gm}{\gamma}
\newcommand{\th}{\theta}
\newcommand{\sq}{\sqrt}
\newcommand{\FNL}{F_{NL}(1,2)}

\newcommand{\nn}{\nonumber}
\newcommand{\bc}{\begin{center}}
\newcommand{\ec}{\end{center}}
\newcommand{\be}{\begin{equation}}
\newcommand{\ee}{\end{equation}}
\newcommand{\ba}{\begin{eqnarray}}
\newcommand{\ea}{\end{eqnarray}}

\newpage
\pagestyle{plain}
\pagenumbering{arabic}

\noindent {\bf  I. INTRODUCTION} \\
The phase ordering dynamics of systems quenched from the disordered
phase to the ordered phase has been extensively studied
\cite{REVIEWS}.
There is a general consensus that at the late stages of domain coarsening
these systems enter a scaling regime \cite{SCALING}, in which the
equal-time, two-point correlation function has the scaling form
\be
   \crt \eq \left<\p({\bf  x},t)\p({\bf  x}+{\bf  r},t)\right>
                = f(r/\lt) \ ,
                \label{C}
\ee
where $\p$ is the scalar order-parameter field, $\lt$ is the
characteristic length scale at time $t$ after the the quench,
$f$ is a scaling function, and angled brackets
indicate an average over initial conditions (and thermal noise, if
present).

A first principles calculation of the scaling function has proved to be
a most difficult task. Even for the simplest model dynamics, that
of a nonconserved order parameter (model A) \cite{HH} with purely
short-ranged (SR) interactions, exact results are rare and available
only for cases of limited physical interest \cite{SOLUTIONS}.

In the past few years closed-approximation schemes for  the
two-point correlation function of the SR model A (SRMA) have been
proposed by a number of authors \cite{WALL.DYN}-\cite{BH93}, based on
a mapping $\prt=\p (\mrt)$ between the order-parameter and an auxiliary
field $\mrt$ which has, near a domain wall, the physical interpretation
of a coordinate normal to the wall.
With this new variable the problem of describing the field at each
instant of time is transformed into a problem of describing the
evolution and statistics of the wall network. This approach enables the use
of a physically plausible and mathematically convenient gaussian distribution
for $m$. Such a distribution is unacceptable for the order parameter field
itself, since this is effectively discontinuous at the
domain size scale.

The application of this sort of approach to both non-conserved and
conserved (model B) dynamics, with purely SR interactions, has recently
received a critical review by Yeung et al. \cite{CLOSE.TH}.
Methods based on a description of the wall dynamics lead to an
approximate {\em linear} equation for $\mrt$, or for its
correlator $\left<m({\bf x},t)m({\bf x}+{\bf r},t)\right>$, \cite{WALL.DYN}.
A different and promising approach, due to Mazenko \cite{MAZ.A}, aims
at deriving a closed {\em non-linear} equation for $\crt$, built on the
equation of motion for model A, using the single assumption that the
field $m$ is gaussian distributed at all times.
It has the advantage of yielding results with a non-trivial dependence
on the spatial dimension $d$ and is also easily extensible to $O(n)$
component systems with topological defects, i.e. with $n\leq d$
\cite{VECTOR}.

The only uncontrolled feature of this approach is the gaussian
assumption. Recent simulation tests have shown, however, that this
assumption is not entirely satisfactory: Blundell et al. \cite{GAUSS.TEST}
have made an absolute test (free of adjustable parameters) of the relation
between two different scaling functions, revealing a disappointing agreement
with the theory. The discrepancy decreases, however, in higher dimensions,
in agreement with an argument \cite{BH93} that the gaussian approximation
becomes exact in the limit $d \to \infty$.  Yeung et al.\ \cite{CLOSE.TH},
using data of Shinozaki and Oono \cite{SO} for $d=3$,
have checked the single-point probability distribution for $m$,
finding it to be flatter at the origin than a gaussian.
It is not difficult to derive an analytical expression for the
{\em two-point} distribution $P(m(1),m(2))$, valid for
$m(1)$, $m(2)$ and $|{\bf  r}|$ small compared to $L(t)$ \cite{P(m)}.
It differs from a gaussian for fixed spatial dimension $d$, but is
consistent with a gaussian in the limit $d\to \infty$.

Despite these reservations, the gaussian approach has been shown to give
good results for the SRMA, displaying most of the expected physical
properties \cite{MAZ.A,VECTOR}.
In this paper we try to extend the limits of this approach by
applying it to model A dynamics with attractive long-range (LR)
interactions.
This application addresses a basic difficulty, not necessarily caused
by the use of the gaussian assumption:
the attempts to extend the approach beyond the simplicity of the SRMA
produce equations for $\crt$ which do not seem to respect the expected
growth-law for the typical domain size $\lt$.
A lack of a proper scaling of the terms in the equation for $C({\bf  r},t)$,
derived naively, is apparent for the case of a scalar order parameter,
namely for the LR model A (LRMA)
and for the SR model B (SRMB), although not for a vector order-parameter
in which case a `naive' dimensional analysis of the equation agrees with
the known growth law (with logarithmic corrections for $n=2$) \cite{BR}.
We shall see how this situation arises for the LRMA and present our
understanding of it. In the case of the SRMB, a naive application of the
method, however, omits the important bulk diffusion process which plays
a vital role in the coarsening. Mazenko \cite{MAZ.B} has
attempted to solve the problem by accounting explicitly for the
bulk diffusion. It is not clear, however, that any analogous mechanism
is present here.

The scalar case, which is usually the more interesting one in the
applications, is exceptional because an extra length, time independent
at late stages, the domain-wall thickness, plays a role in the dynamics,
and therefore power counting of lengths by dimensional analysis may not
yield the right scaling (in terms of the characteristic length) for the
different parts in the equation of motion.
For the SRMB \cite{MODEL.B} and the LRMA \cite{BR,GROWTH} dynamics the
growth laws are $\lt \sim t^{1/3}$ and $\lt \sim \tuAoAPsg$ (for $\sgLA$),
respectively, for $n=1$, and $\lt\sim t^{1/4}$ and $\lt \sim \tuAosg$,
respectively, for $n>2$ (with logarithmic corrections for $n=2$ \cite{BR}),
where $\ZLsgLB$ is the exponent describing
the LR interactions, which decay as $1/r^{d+\sigma}$. For $n=1$ and
$\ALsgLB$, the long-range interactions are irrelevant and the growth-law
is the same as for the SR case \cite{BR,GROWTH}.
The SRMA, however, is exceptional since the predicted growth-law,
$\lt \sim \tuH$, is the same for both the scalar and
vector order-parameters, accidently allowing for a 'naive' dimensional
analysis of the scalar equation of motion to agree with the growth-law.
In this case the role of the extra length in the scalar equation can be
ignored as the result of two canceling errors \cite{GROWTH}.
Therefore we wonder if the success of the Mazenko method with this scalar
model might be somewhat fortuitous. In other words, we raise the question
of whether this approach (or any other closed theory), naively applied,
can succeed for those dynamical models where naive dimensional analysis
gives the wrong growth law. In this respect it is interesting that
a straightforward application of the method of Kawasaki, Yalabik and
Gunton \cite{KYG} to the LRMA \cite{HRT} also gives the wrong growth
law for $n=1$, i.e.\ it gives the $t^{1/\sg}$ growth suggested by
naive dimensional analysis.

In this paper we have developed an extension of Mazenko's approach for
the LRMA. Besides having interest by its own right, the study of this
model provides a test-bed for the approach and shows some of the
difficulties any approximate closed theory must resolve.

\bigskip

\noindent {\bf  II. THE MODEL WITH LONG-RANGE INTERACTIONS} \\
We consider a system with long-ranged {\em attractive} interactions,
falling off with distance as $r^{-(d+\sg)}$. A suitable Hamiltonian
functional of the scalar field is
\be
   H[\p] = \intd r [(\nabla \p)^2/2 + \vp]
         + (\jlr/2)\intd r \intd r' [\pr -\pR]^2/\kdrR \ ,
                \label{Hamiltonian}
\ee
where as usual we have taken the short-range part to have the
Ginzburg-Landau form, $\jlr>0$, and $\vp$ has a local maximum at $\p=0$
and global minima at $\p=\pm 1$. The model is well defined for
$\ZLsgLB$.
The equation of motion for a non-conserved field reads
$\partial \phi/\partial t = -\delta H/\delta \phi$, i.e.\
\be
   \partprt = \nabBp - \Dvp + \Dvlrprt \ ,
                \label{eq.phi}
\ee
where $\Dvp=dV/d\p$ and the LR force is given, both in real and Fourier
space, as
\ba
   \Dvlrprt & = & \jlr \intd r' [\pR - \pr]/\kdrR   \\
            & = & \jlr\,\hds \intdk \ \pk \ k^{\sg} \exprk \ ,
                \label{VLR}
\ea
and
\ba
   \hds & = & \Qds \ \frac{\sqrt{\pi}}{2^{\sg}}
         \frac{\Gamma(-\frac{\sg}{2})}{\Gamma(\frac{1+\sg}{2})}\ ,
                 \label{h}   \\
   \Qds & = & \pi^{\frac{d-1}{2}}
         \frac{\Gamma(\frac{1+\sg}{2})}{\Gamma(\frac{d+\sg}{2})} \ .
                \label{q}
\ea
In (\ref{eq.phi}) noise is absent since temperature is an irrelevant
variable \cite{NOISE}.
{}From an analysis of (\ref{eq.phi}), assuming the validity of the scaling
hypothesis (1), the following growth-law has been predicted
for a scalar order-parameter \cite{BR,GROWTH}
\ba
   \lt & \sim & \tuAoAPsg \ , \ \ \ \ZLsgLA \nn \\
       & \sim & \tuH  \ , \ \ \ \ALsgLB \ ,
                \label{growth.law}
\ea
in which the crossover $\sg=1$ separates the regime where domain
growth is faster due to the LR correlations from the regime where
these become irrelevant \cite{BR,GROWTH,CARDY}.

\bigskip

\noindent {\bf  III. THE SCALING EQUATION} \\
To obtain an equation for the two-point correlation function (\ref{C})
we multiply (\ref{eq.phi}), evaluated at point $(1)\eq ({\bf r}_1,t_1)$,
by $\p$ evaluated at point $(2)\eq ({\bf  r}_2,t_2)$ and average over the
ensemble of initial conditions yielding, at equal-times,
\be
 \Half \partcAB = \nabBcAB - \left<\p(2)\DvpA \right>
                           + \left<\p(2)\DvlrpA \right> \ .
                \label{eq.C}
\ee
We will call $\left<\p(2)\DvpA \right>$ and $\left<\p(2)\DvlrpA\right>$
the `non-linear' (NL) and the `long-range' (LR) terms of the equation
for $\crt$, where ${\bf r}={\bf r}_1-{\bf r}_2$.
In (\ref{eq.C}) the LR term reads, both in real and Fourier space,
\ba
   \left<\p(2)\DvlrpA \right> & = & \jlr \intd r'\,[\cRt -\crt]/\kdrR \\
                       & = & \jlr\,\hds \intdk\,\ck \ k^{\sg} \exprk \ ,
                \label{FCLR}
\ea

Assuming the existence of a late-time single-scaling regime, we
expect $\crt$ to take the scaling form (\ref{C}), in terms of
which (\ref{eq.C}) reads
\be
 -\Half \Dlol \ x f' = \AoluB (f''+ \dMAox f')
                     - \left<\p(2)\DvpA \right>
                     + \left<\p(2)\DvlrpA \right> \ ,
                \label{eq.f}
\ee
where $x=r/\lt$ is the scaling variable and $f'=df/dx$, etc.
In the equation above $\dot{L}/L \sim 1/t$, if $\lt$ grows as a
power-law.
The LR term now reads
\ba
\left<\p(2)\DvlrpA \right> & = & \jolusg \intd x'\,[f(x')-f(x)]/\kdxX \\
                   & = & \jholusg \intdy \ g(y) \ y^{\sg} \expxy \ ,
                \label{FfLR}
\ea
where $g(y)$ is the Fourier transform of $f(x)$, and $y=k\lt$.

{}From an analysis of (\ref{eq.phi}) for an isolated, stationary, planar wall,
we find that to leading order the equilibrium {\em planar} wall profile
saturates as
\be
   1-\p^2(r) \sim \frac{\jlr}{V_0'' \ r^{\sg}} \ , \ (r\to \infty) \ ,
                \label{profile}
\ee
where $V_0''= (d^2V/d\p^2)_{\p^2=1}$ and $r$ is the distance from the wall.
Hence we expect that throughout the bulk region $|\p|$ will be
below saturation by an amount $\sim 1/\ltusg$.
Even with this power-law decay we still expect there to be well-defined
walls, with a time-independent `thickness' $w$, defined for example from
(15) via $w^{\sigma} = J_{LR}/V''_0$. Therefore, domain walls may be
regarded as `sharp' at late-times, when $L(t) \gg w$.
It follows that Porod's law \cite{POROD}, $g(y) \sim \Ads/y^{d+1}$ for
$y\gg 1$, holds within the regime $kw \ll 1 \ll kL(t) \equiv y$
(corresponding to $w \ll r \ll L(t)$ in real space), in which case eq.
(\ref{FfLR}) yields for the leading scaling behaviour of the LR term, as
$x\to 0$,
\ba
   \left<\p(2)\DvlrpA \right> \!\!
   & = & \!\! \jholusg \left( \intdy\,g(y)\,y^{\sg} \! +
         \frac{\Ads\,\hdAMs}{\Bpiud}\,\xuAMsg \! + ...
         \right) , \ \ZLsgLA \nn \\
   & = & \!\! \jholusg \left( \frac{\Ads\,\hdAMs}{\Bpiud} \AoxusgMA
         + O(1)     \right) , \ \ALsgLB \ .
                \label{LR.smallr}
\ea
This result will be exploited below to determine the amplitude
$A(d,\sigma)$ of the Porod tail, within the gaussian approximation.

\bigskip

\noindent {\bf  IV. THE GAUSSIAN APPROXIMATION} \\
In order to transform (\ref{eq.C}) or (\ref{eq.f}) into a closed
equation we need to express the NL term as some approximate non-linear
function of $\crt$.
A key idea, exploited by several authors [6-11] within SR
model A dynamics, is to employ a non-linear mapping between the order
parameter $\prt$, which at the scale of $\lt$ is effectively
discontinuous near walls, and an auxiliary `smooth' field $\mrt$, whose
zeros define the wall network. This introduces the wall structure
into the problem and allows the approximation to be implemented through
the new field.

{}From the equation of motion (\ref{eq.phi}) we can see that, just like in the
SRMA, if the initial field satisfies $|\p|\leq 1$ then this condition
will hold at all times, assuring that a one-to-one mapping can be
defined.
For this model we have in mind, following Mazenko's treatment for SR
interactions, to identify the field $\mrt$ at points ${\bf r}$ near
domain walls as {\em the (signed) distance to the nearest wall}
(along its local normal), with the sign of $m$ being that of $\phi$.
This determines $m$ uniquely when $m \ll L(t)$. To specify $m$ everywhere
in space, we define the function $\prt=\p(\mrt)$ by extending Mazenko's
suggestion \cite{MAZ.A} of using the equilibrium {\em planar} domain
wall profile function for an isolated wall, with $m$ the coordinate
normal to the wall, i.e. the function $\phi(m)$
is specified by the equation
\be
   0 = \DDp - V'(\p(m)) +  \jlr \intdMA y \intinf dm'
       \frac{[\p(m')-\p(m)]}{[(m'-m)^2+{\bf y}^2]^{\frac{d+\sg}{2}}} \ ,
                \label{V'm}
\ee
with boundary conditions $\p(0)=0$ and $\p(m)\to {\rm sign}\,(m)$ for
$|m|\to \infty$. Using (\ref{V'm}) we rewrite the NL term in
(\ref{eq.C})-(\ref{eq.f}) as
\ba
 \left<\p(2)\DvpA\right> \! & = & \! \left<\p(2)\DDpA\right> \ + \nn \\
                 &   &  \! \Qds \intinf ds \Aoks
             \left<\p (\mB )[\p (\mA +s)-\p(\mA )]\right> \\
                 & = & \! \intk\!\intK \tilde{\p}(k) \tilde{\p}(k')
             \left[\jhds k^{\sg}\!-\! k^2 \right]
             \left<\expmk \right> \ ,
                \label{NL}
\ea
where $\hds$ and $\Qds$ are given by (\ref{h})-(\ref{q}), and
$\tilde{\phi}(k)$ is the Fourier transform of $\phi(m)$.

Following Mazenko \cite{MAZ.A}, we now make the key assumption that
$\mrt$ is a gaussian field (with zero mean) at all times, with a pair
distribution function
\be
    P(\mA ,\mB ) = N \ \exp \left[ - \frac{1}{2(1-\gm^2)}
    \left( \frac{\mA^2}{\SA} + \frac{\mB^2}{\SB}
         - \frac{2\gm\mA\mB}{\sq{\SAB}} \right) \right]  \ ,
                \label{P2}
\ee
\be
   \SA=\left<\mA^2\right>,\ \ \gm(1,2)=\frac{\left<\mA\mB\right>}
   {\sq{\SAB}}, \ \ N=\frac{1}{2\pi\sq{(1-\gm^2)\SAB}} \ .
                \label{moments}
\ee
We also note that, as the walls become effectively sharp in the late-time
regime, we can use the profile $\p(m)={\rm sign}\,(m)$ to evaluate the
leading contribution to the scaling functions.
{}From (\ref{profile}) we expect the effect of ignoring the power tail
in the profile is to neglect a quantity of relative order
$\sim 1/\lt^{\sg}$ in the LR part of (\ref{NL}).
The purely SR part of the NL term is then simply given, as a non-linear
function of $\crt$, by Mazenko's result for the SRMA \cite{MAZ.A}
\be
  \left<\p(2)\DDpA \right> = -\frac{2}{\pi\SA}\tan\left(\pic\right) \ .
                \label{SRNL}
\ee

Deriving a similar result for the LR part of the NL term is more tricky.
There are three different ways to perform the calculation: we will
outline the basic steps of each one.
Representing $\p(m)$ in Fourier space and Taylor expanding $\p(m\!+\!s)$
in powers of $s$, using the gaussian property and returning to real
space, gives the formal expansion \cite{comparison}
\ba
 \FNL  & \eq &  \Qds \intinf ds \Aoks
              \left< \p (\mB )[\p (\mA +s)-\p(\mA )] \right> \nn \\
       & = &  \Qds \intinf ds \Aoks
              \sum_{n=1}^{\infty} \frac{s^{2n}}{(2n)!}
              2^n \frac{\partial^nC(1,2)}{\partial \SA^n} \ .
                \label{LRNL1}
\ea
Using $C(1,2) = \left<{\rm sign}\,(\mA){\rm sign}\,(\mB)\right>$, the
integral representation ${\rm sign}\,(m) = 1/(i\pi) \intinf dz \exp[izm]/z$,
and the gaussian property, the series can be summed.
Finally, differentiating with respect to $C_0(1,2)=\gm \sq{\SAB}$,
performing the $z$ and $s$-integrals and integrating back, yields the
non-linear function
\be
   \FNL = \frac{\jads}{\SA^{\sg/2}}
          \int^{\pic}_0 d\th \sec^{\sg}(\th) \nn \\ \ ,
                \label{LRNL}
\ee
\be
   \ads =\hds \ \frac{2^{1+\sg/2}\Gamma(\frac{1+\sg}{2})}{\pi^{3/2}} \ .
                \label{A}
\ee
Alternatively, we can take $\p(m)={\rm sign}\,(m)$ from the start and do
the $s$-integral, giving
\be
  \FNL = -\frac{2\jlr Q(d,\sigma)}{\sg}
      \left<\frac{{\rm sign}\,(\mA )\,{\rm sign}\,(\mB )}
{|\mA |^{\sg }}\right> \ ,
\label{LRNL2}
\ee
use integral representations for ${\rm sign}\,m$ and $1/|m|^{\sg}$,
do the gaussian integral, differentiate with respect to $C_0(1,2)$,
perform the remaining integrals, and finally integrate back over $C_0(12)$,
yielding the same result.
Taking into account (\ref{profile}) and using a ${\rm sign}\,m$ profile,
(\ref{LRNL2}) can be recognized as the leading order result for a
$\p^4$-potential NL term, i.e. $<\!\p(2)\p(1)(1-\p^2(1))\!>$
\cite{delta}.
Finally, the simplest derivation is to take the integral representation
of ${\rm sign}\,m$ in (\ref{LRNL2}), use the gaussian
property, differentiate with respect to $C_0(1,2)$ and do the gaussian
integral,
leading to the same point as the first calculation before its final
integrations. This derivation, however, does not provide the appealing
intermediate expressions (\ref{LRNL1}) and (\ref{LRNL2}).

According to our identification of $\mrt$ as a distance from the interface,
we expect $S_0 \equiv \langle m^2 \rangle$ to have the
scaling form $S_0 = \ltuB$, which can be used along with (\ref{SRNL})
and (\ref{LRNL}) to rewrite equation (\ref{eq.f}) for the scaling
function in the form
\ba
 &   & -\Half \Dlol \ x f' \nn =
 \AoluB \left(f''+\dMAox f'+\Bopi \tan\left(\pif\right) \right) \nn \\
 & + & \jolusg \left( \intd x'\frac{[f(x')-f(x)]}{\kdxX}
                 - \ads \int^{\pif}_0 d\th \sec^{\sg}(\th) \right) \ ,
                \label{finaleq.f}
\ea
For $\sg <2$ the SR part in (\ref{finaleq.f}), scaling as $1/L^2$, is
negligible compared to the LR part, scaling as $1/L^{\sg}$, and can be
ignored (but see the discussion in section VI!). Demanding that the
left side of (\ref{finaleq.f}) balance the terms of order $1/L^\sigma$
on the right requires $\dot{L}/L \sim 1/L^{\sigma}$, i.e.\ that
$L(t) \sim t^{1/\sigma}$. Note that this disagrees with the expected
form (\ref{growth.law})! In section V we will argue that a resolution
of this discrepancy requires us to drop the left side of ({\ref{finaleq.f})
in leading order. For the moment, however, we pursue the original (and
{\em a priori} natural) assumption that the left side scales as $1/L^\sigma$
and write
\be
    \lt = \left( \jlr \, \mu \, t \right)^{1/\sg} \ ,
                \label{L}
\ee
where $\mu$ is to be determined. Dropping the SR terms from (\ref{finaleq.f})
gives the final equation for the scaling function $f(x)$:
\be
0 = (\mu/2\sigma)\,xf' + \intd x'\frac{[f(x')-f(x)]}{\kdxX}
- \ads \int^{\pif}_0 d\th \sec^{\sg}(\th) \ .
\label{scaling}
\ee

Equation (\ref{scaling}) has to be solved numerically for general scaling
variable $x$. However, it is straightforward to derive analytically the
behaviour for small and large $x$.
Using the Porod's law form  $f(x)=1-\alds x +...$ for small $x$ (it is simple
to show that this is only consistent short-distance behavior),
we find that the LR part of the NL term, (\ref{LRNL}), has
a leading scaling behaviour as $x\to 0$ which is similar to
(\ref{LR.smallr})
\ba
   \FNL & = &
    \jaolusg\left(\frac{B(\frac{\AMs}{2},\frac{\AMs}{2})}{2^{1+\sg}}
   -\left(\frac{\pi\al}{2}\right)^{\AMs}\frac{\xuAMsg}{\AMs}+ ...
    \right) \ , \ \ZLsgLA \nn \\
   & = & \jaolusg\left( \left(\frac{\pi\al}{2}\right)^{\AMs}
         \frac{1}{\xusgMA(\sMA)} + O(1)\right) \ , \ \ALsgLB \ ,
                \label{LRNL.smallr}
\ea
where $B(x,y)$ is the beta function.
Performing a small-$x$ expansion of eq. (\ref{scaling}),
we find that the dominant terms for $x \to 0$ are obtained from the
terms multiplying $J_{LR}$ in (27), whose small-$x$ expansions are given
by (\ref{LR.smallr}) and (\ref{LRNL.smallr}). Matching powers of $x$
for general $\ZLsgLB$, and using $\Ads = - \alds (2\pi)^d /h(d,1)$ in
(\ref{LR.smallr}) (which follows from Fourier transforming the Porod
tail \cite{AMPLITUDES}), we find
\be
   \alds = \frac{\sq{2}}{\pi} \ \left(\frac{\Gamma(\frac{d+1-\sg}{2})}
         {\Gamma(\frac{d+1}{2})} \right)^{1/\sg} \ , \ 0<\sigma<2 \ ,
                \label{alpha}
\ee
for the coefficient of $x$ in the small-$x$ expansion of $f(x)$ \cite{alpha}.
For $\sigma=2$, this reduces to the SR result $\al(d,2)=2/(\pi\sq{d-1})$.

For $\sigma<1$, (\ref{alpha}) was obtained by matching the terms of
$O(x^{1-\sigma})$ in (\ref{LR.smallr}) and (\ref{LRNL.smallr}). The
leading (constant terms) yield an interesting sum rule to be satisfied
by the structure factor scaling function $g(y)$ for this range of $\sigma$:
\be
   \intdy \ g(y) \ y^{\sg} = \frac{2^{\sg/2}}
 {\Gamma(\frac{2-\sg}{2}) \ \sin(\frac{1+\sg}{2}\pi)} \ , \ 0<\sigma<1\ .
                \label{LR0}
\ee

We now look at the large-$x$ asymptotic form of eq. (\ref{scaling}),
and discuss the large-$x$ behaviour of $f(x)$. In this limit,
$f(x) \to 0$ and the final two terms in (\ref{scaling}) become
$g(0)/x^{d+\sg}$ and $-a(d,\sg)\,\pi f(x)/2$ respectively.
In this regime, (29) can be integrated to give
\be
f(x) \to \frac{2\sigma g(0)}{[(d+\sg)\mu - \pi|a|]}\,
\frac{1}{x^{d+\sigma}} + \frac{A}{x^{|a|\pi\sigma/\mu}}\ ,
\ee
where we note from (25) and (6) that $a$ is negative. In general, both
terms in (33) will be present in the large-$x$ solution. On physical
grounds, however, we do not expect $f(x)$ to fall off with distance
more slowly (in a power-law sense) than the underlying interactions,
which decay as $r^{-(d+\sigma)}$.
(An exception is when sufficiently long-range
power-law spatial correlations are present in the $t=0$ state. This
power-law can then persist for general times \cite{BHN}. Here, however,
we consider only short-range correlations in the initial state.)
We infer that either $\mu < |a|\pi\sg/(d+\sigma)$, so that the second
term in (33) is subdominant for large $x$, or $A = 0$. The first
possibility, however, implies that the coefficient of the (dominant)
first term in (33) is negative (since $g(0) > 0$ by definition),
i.e.\ $f(x)$ approaches zero from below,
which also seems unphysical (and disagrees with numerical simulations
\cite{HAY2}). We conclude that the only physically sensible
possibility is that $A$ vanishes in (33). This can, presumably, only
happen for a special choice of $\mu$, so the condition $A=0$ determines
$\mu$. This mechanism is very similar to that which determines $\mu$
for short-range interactions \cite{MAZ.A,VECTOR}. Note that, if $f(x)$
is to approach zero from above for $x \to \infty$, (33) gives the inequality
\begin{equation}
\mu > \pi |a|/(d+\sigma)\ .
\end{equation}
A sum rule for $\mu$ can be obtained by integrating (29) over
space:
$$
\mu = \frac{2\sigma |a|}{dg(0)} \int d^dx \int_0^{\pi f/2} d\theta\,
\sec^{\sg}\theta\ .
$$

Finally, it should be noted that the above analysis implicitly assumes
that $g(0)$ is finite, i.e.\ that $f(x)$ decays faster than $x^{-d}$.
In fact, the mathematical structure allows for $f(x) \sim x^{-p}$
with $p<d$ \cite{Joao}, but we reject this possibility on the
physical grounds that we appealed to before, namely that, at least
for initial states with only short-ranged spatial correlations, the
scaling function should not decay with a smaller power than the underlying
interactions.

\bigskip

\noindent {\bf  V. TWO-TIME CORRELATIONS} \\
The gaussian approach can also be used to evaluate the two-time
correlation function
$C({\bf  r},t_1,t_2) = \langle \phi({\bf  x},t_1)\phi({\bf  x}+{\bf  r},t_2)
\rangle$ and, in particular, the autocorrelation function
$A(t_1,t_2) = C(0,t_1,t_2)$. The calculation is simplest in the limit
$t_2 \gg t_1$, when $C \to 0$ and the full nonlinear equation can be
linearized, Fourier transformed, and explicitly integrated. In this
regime the analog of (9) for two-time correlations reads, in Fourier
space (dropping the SR term on the right),
\begin{equation}
\frac{\partial C_{\bf  k}}{\partial t_2} = -J_{LR}|h|k^\sg C_{\bf  k}
                     + \frac{\pi|a|}{2\mu t} C_{\bf  k}\ ,\nonumber
\end{equation}
where (28) has been used for $L(t)$. We integrate (35) forward
from time $\alpha t_1$, where $\alpha \gg 1$ ensures that the condition
$t_2 \gg t_1$, required for the validity of (35), holds at all times.
This gives
\begin{equation}
C_{\bf  k}(t_1,t_2) = C_{\bf  k}(t_1,\alpha t_1)\,
                     \left(\frac{t_2}{\alpha t_1}\right)^{\pi|a|/2\mu}\,
                     \exp\{-J_{LR}|h|k^\sg (t_2 - \alpha t_1)\}\ .
\nonumber
\end{equation}
Using the scaling form $C_{\bf  k}(t_1,\alpha t_1) = L_1^d g_\alpha(kL_1)$,
where $L_1=L(t_1)$, and summing over ${\bf  k}$ for $t_2 \gg \alpha t_1$
gives the autocorrelation function
\begin{equation}
A(t_1,t_2) = {\rm const}\,(L_1/L_2)^{d-\pi|a|\sg/2\mu}\ ,
\end{equation}
where `const' is clearly independent of $\alpha$. The physical requirement
that $A$ {\em decrease} with increasing $t_2$ gives the inequality
$\mu > \pi|a|\sg/2d$, which is guaranteed by (34) for $d > \sigma$.

The connection between the parameter $\mu$ and the exponent describing
the decay of the autocorrelation function is similar to that obtained
for purely short-ranged interactions \cite{VECTOR,LM}.

\bigskip

\noindent{\bf  VI. DISCUSSION AND SUMMARY} \\
We have extended the original Mazenko gaussian approach \cite{MAZ.A} to
the LR model and evaluated the late-time leading contribution to the NL
term of eq. (\ref{finaleq.f}), yielding a dominant LR part given by
(\ref{LRNL}), which is of order $1/L^{\sg}$.
An infinitely sharp wall profile has been used which amounts to
neglecting a quantity of relative order $1/L^\sg$.
The LR term in the equation, (\ref{FCLR})-(\ref{FfLR}), is of the same
order and has an amplitude which is a function of $x$, $d$ and $\sg$,
but its non-local nature (i.e. its dependence on the values of $f(x)$
everywhere) makes the problem particularly hard to handle.

Despite the profile power-law decay (\ref{profile}) induced by the LR
interactions, the scaling function exhibits Porod's law, i.e.\ a linear
short-distance behaviour in real space with coefficient given by
(\ref{alpha}). This is consistent with the assumption that at late-times
there are well-defined walls with a constant `width' independent of $L(t)$.
This is an important point of principle, on which the identification of
the field $\mrt$ and the mapping (\ref{V'm}) rely, and also a key ingredient
in the first-principles derivation \cite{BR,GROWTH} of the growth-law
(\ref{growth.law}).

The central question we want to address in this paper is whether the
gaussian theory is able to yield the correct growth-law for this
model. We have seen that the `naive' application of the gaussian approach
present in section IV ostensibly gives the wrong growth law: (\ref{L})
instead of (\ref{growth.law}).
A related problem is the SRMB to which Mazenko has attempted to apply
the gaussian approach \cite{MAZ.B}, yet the correct growth-law does not
come out of the theory as cleanly as in the SRMA. In this system local
conservation imposes a bulk diffusion process which controls the
interface motion and delays domain coarsening relative to the purely
relaxational dynamics of model A. There are some common features
between the dynamics of a conserved and a LR interacting field,
namely the existence of a bulk profile which
relaxes rapidly to a non-saturating value as the walls move. One key
difference, though, is that the true growth law for SRMB ($t^{1/3}$)
is {\em faster} than that obtained by a naive application of the
gaussian approach (without allowing for bulk diffusion), which gives
$t^{1/4}$.

Before implementing any approximation we focus the analysis on the
exact equation (\ref{eq.f}).
If the growth-law (\ref{growth.law}) holds, the time-derivative term
must be negligible compared to the LR term (\ref{FfLR}), which scales
as $L^{-\sg}$, and therefore the NL term must have a leading contribution
of order $1/L^{\sg}$ which exactly cancels the LR term in the scaling limit.
In fact, this condition determines the late-time leading contribution to
the scaling function. Within the gaussian approximation, it amounts to
neglecting the first term in (\ref{scaling}) (which came from the
left-hand side of (\ref{finaleq.f})), to give
\be
0  =  \intd x'\frac{[f(x')-f(x)]}{\kdxX}
- \ads \int^{\pif}_0 d\th \sec^{\sg}(\th)\ .
\label{scaling1}
\ee
Solving this equation gives the scaling function $f(x)$, within the gaussian
approach, provided the growth law is {\em slower} that $t^{1/\sg}$. However,
there seems to be no way to {\em determine} the growth law within this scheme.
Moreover, (\ref{scaling1}) has a serious shortcoming. If we integrate the
equation over all space, the first term drops out, giving the sum rule
\be
\int_0^\infty dx\,x^{d-1} \int_0^{\pi f(x)/2} d\theta\,\sec^\sg(\theta) = 0\ .
\label{sumrule}
\ee
Since the integrand is positive definite, the only way this sum rule can
be satisfied is for $f(x)$ to be negative for some range (or ranges) of $x$,
with sufficient negative weight to satisfy (\ref{sumrule}). This seems
{\em a priori} improbable for a nonconserved order parameter, and indeed
numerical simulations \cite{HAY2} show no hint of it.

We emphasize, however, that since our fundamental equation (\ref{eq.f})
is {\em exact}, the analogue of (36) obtained {\em without making the
gaussian approximation} must be exactly true. Because the true growth is
{\em slower} than $t^{1/\sigma}$, the left side of (12)
is negligible in the scaling limit. Taking the
Fourier transform of the equation, and setting $k=0$, the `long-range'
term vanishes. This leaves the identity
\be
    \intd x \left<\p(2)\DvpA \right> = 0 \ ,
                \label{cancelation}
\ee
of which (39) is the special case obtained within the gaussian
approximation.

Our results seem to indicate that the gaussian approach, applied to the
bulk equation of motion, is unable to account for the qualitative
feature of coarsening in systems with long-range interactions.
However, we cannot make a definitive statement as we have not exhausted
all the possible choices for the gaussian field. There may exist a
mapping definition which is physically more appropriate and works better
then (\ref{V'm}).
What seem to be clear is that, beyond the simple nonconserved system with
short-range interactions, one cannot apply the gaussian approach in a
straightforward and `naive' manner to construct a closed equation for the
scaling function. Just as for the conserved scalar system,
a deeper understanding of the underlying physics may be required in order
to implement a more controlled approximate scheme.
It is possible that this might be achieved by means of an
interfacial approach \cite{HAY3}.

To summarize, the gaussian approach, naively applied,
is unable to yield a growth-law different from that obtained from
dimensional analysis of the linear terms in the equation of motion.
As mentioned above, the failure of the gaussian approach in this
context could be due to the particular choice employed for the
mapping between $\p$ and the gaussian field. For example, one
can in principle use the same mapping as Mazenko \cite{MAZ.A},
$\p''(m)=V'(\p(m))$, appropriate to a purely short-range interactions.
However, this leads to an inconsistent scaling analysis of the equation
for $\crt$ (e.g. at short-distances there is no LR part in the NL term
to match the LR term (\ref{LR.smallr})), and $\mrt$ can no longer be
regarded as a distance from a wall. By contrast, the mapping employed
here, defined by eq. (\ref{V'm}), seems far more natural and physically
suitable for a system with LR interactions.

Finally we note that the present methods can also be used for a vector
order parameter with long-range interactions. In that case the
$t^{1/\sg}$ growth obtained within the gaussian approach is correct
(apart from logarithmic corrections for $n=2$ \cite{BR}). The purpose
of the present paper, however, is to test the method on those systems
which provide the greatest challenge, i.e.\ scalar systems, in the
hope that the difficulties identified here may stimulate the
development of more robust approximation schemes.

\bigskip

\noindent {\bf  ACKNOWLEDGEMENTS} \\
A. B. thanks T. Ohta for a stimulating discussion.
J. F. thanks JNICT (Portugal) for financial support.


\begin{thebibliography}{99}
\bibitem{delta}
This shows that, in contrast to the short-range case, $\p(m)(1-\p^2(m))$
cannot be treated as a representation for the derivative of a delta function
(its normalization would be infinite).
\end{thebibliography}
\end{document}